\documentclass[pdflatex,sn-nature]{sn-jnl}% Style for submissions to Nature Portfolio journals
\pdfoutput=1
\usepackage{graphicx}%
\usepackage{multirow}%
\usepackage{amsmath,amssymb,amsfonts}%
\usepackage{amsthm}%
\usepackage{mathrsfs}%
\usepackage[title]{appendix}%
\usepackage{xcolor}%
\usepackage{textcomp}%
\usepackage{manyfoot}%
\usepackage{booktabs}%
\usepackage{algorithm}%
\usepackage{algorithmicx}%
\usepackage{algpseudocode}%
\usepackage{listings}%
\usepackage{float}%
\usepackage{natbib}%
\usepackage{lineno}
\usepackage{bm}
\theoremstyle{thmstyleone}%
\theoremstyle{thmstyletwo}%

\theoremstyle{thmstylethree}%

\begin{document}
%\linenumbers%

\title[Article Title]{Laboratory formation of scaled astrophysical outflows}


\author[1]{\fnm{Shun-yi} \sur{Yang}}

\author*[1,2]{\fnm{Guang-yue} \sur{Hu} }\email{gyhu@ustc.edu.cn}

\author[1]{ \fnm{Chao} \sur{Xiong}}

\author[1]{ \fnm{Tian-yi} \sur{Li}}

\author[1]{\fnm{Xue-cheng} \sur{Li} }

\author*[1]{ \fnm{Hui-bo} \sur{Tang}}\email{tanghb@ustc.edu.cn}

\author[1]{ \fnm{Shuo-ting} \sur{Shao}}

\author[1]{ \fnm{Xiang} \sur{Lv}}

\author[1]{ \fnm{Chen} \sur{Zhang}}

\author[3]{ \fnm{Ming-yang} \sur{Yu}}

\affil[1]{\orgdiv{School of Nuclear Science and Technology, School of Physical Science, and CAS Key Laboratory of Geospace Environment}, \orgname{University of Science and Technology of China}, \orgaddress{\city{Hefei}, \postcode{230026},\country{China}}}

\affil[2]{\orgdiv{CAS Center for Excellence in Ultra-intense Laser Science, Shanghai Institute of Optics and Fine Mechanics}, \orgname{Chinese Academy of Sciences}, \orgaddress{\city{Shanghai}, \postcode{201800}, \country{China}}}

\affil[3]{\orgdiv{Shenzhen Key Laboratory of Ultra-intense Laser and Advanced Material Technology,  Center for Advanced Material Diagnostic Technology, and College of Engineering Physics}, \orgname{Shenzhen Technology University}, \orgaddress{\city{Shenzhen}, \postcode{518118}, \country{China}}}

\abstract{Astrophysical systems exhibit a rich diversity of outflow morphologies, yet their mechanisms and existence conditions remain among the most persistent puzzles in the field. Here we present scaled laboratory experiments based on laser-driven plasma outflow into magnetized ambient gas, which mimic five basic astrophysical outflows regulated by  interstellar medium, namely collimated jets, blocked jets, elliptical bubbles, as well as spherical winds and bubbles. Their morphologies and existence conditions are found to be uniquely determined by the external Alfvénic and sonic Mach numbers $M_{e-a}$ and $M_{e-s}$, i.e. the relative strengths of the outflow ram pressure against the  magnetic/thermal pressures in the interstellar medium, with transitions occurring at $M_{e-a}$ $\sim$ 2 and 0.5, as well as $M_{e-s}$ $\sim$ 1. These results are confirmed by magnetohydrodynamics simulations and should also be verifiable from existing and future astronomical observations. Our findings provide a quantitative framework for understanding astrophysical outflows.}


\maketitle
\newpage
\section*{Introduction}\label{sec1}

Spectacular astrophysical outflows are ubiquitous in the universe and have been observed in protostars (YSOs), planetary nebulae (PNs), supernovae, active galactic nuclei, and quasars, etc. The most common outflow morphologies are collimated jets, blocked jets, and circular, elliptical, and bilobed bubbles/cavities,  as well as quasi-spherical winds.~\cite{bally2016protostellar,balick2002shapes,frank1999bipolar} Understanding their existence condition and behavior is pivotal for unraveling the transport of mass, momentum and energy in celestial evolutions. 

In the standard model of interacting stellar winds ~\cite{kwok1978origin,Balick1987The,icke1988blowing,frank1990stellar}, magneto-centrifugally accelerated, wide-angle, fast stellar winds~\cite{blandford1982hydromagnetic,JFerreira1997,ainsworth2013subarcsecond} are modulated by the thermal ~\cite{matsakos2009two} and/or magnetic pressures~\cite{spruit1997collimation,Matt2003} in the surrounding interstellar medium (ISM). Analytical theories~\cite{frank1996apj,canto1980stellar,canto1988formation,icke1988blowing,frank1999bipolar,carrasco2010magnetized,kastner2012chandra} and numerical simulations~\cite{Icke1992,balick2017models} based on this model can replicate some of the observed outflow morphologies. However, astronomical observations remain undersampled and not well resolved~\cite{zinnecker1998symmetrically,bjerkeli2016resolved}, consequently the effects of parameters such as ISM's thermal and magnetic pressures on the different outflow morphologies have yet to be established.

Laboratory experiments~\cite{remington1999modeling,BLACKMAN2022101661} with controllable and reproducible conditions can be scaled to astrophysical environments and help validating theoretical and numerical models. Recently, substantial efforts using laser or pulsed power facilities have produced jet via inertial or magnetic collimation or radiative cooling~\cite{albertazzi2014laboratory,ciardi2013astrophysics,yurchak2014experimental,higginson2017enhancement,manuel2019magnetized,shigemori2000experiments,revet2021laboratory,hartigan2009laboratory,ciardi2008curved,lebedev2005magnetic,ciardi2007evolution}. While,  the quantitative formation conditions remain unclear and other basic outflow morphologies are still under study. 

Here we report on laboratory generation of scaled astrophysical outflows to quantitatively investigate the effects of the ISM's thermal and magnetic pressures on the astrophysical outflow. Five basic outflow morphologies are demonstrated, namely collimated jets, blocked jets, elliptical bubbles, spherical winds and bubbles. We found that the outflow morphology is uniquely determined by the external Alfvénic  and sonic Mach number $M_{e-a}=V/V_a$ and $M_{e-s}=V/C_s$, where \textit{V} is outflow (or ram) velocity, $V_a$ and $C_s$ are the Alfvén  and sonic velocity of ISM, respectively. Moreover, transitions among the outflow morphologies occur at $M_{e-a}$ $\sim$ 2 and 0.5, as well as $M_{e-s}$ $\sim$ 1. That is, the relative magnitudes of the ram, and the ISM's magnetic and thermal pressures determine the  distinct outflow morphologies. The  quantitative relationship  among the ISM parameters and the outflow morphologies are also confirmed with magnetohydrodynamic (MHD) simulations.
 
\section*{Experimental setup}\label{sec2}

\begin{figure}[h]
\centering
\includegraphics[width=0.8\textwidth]{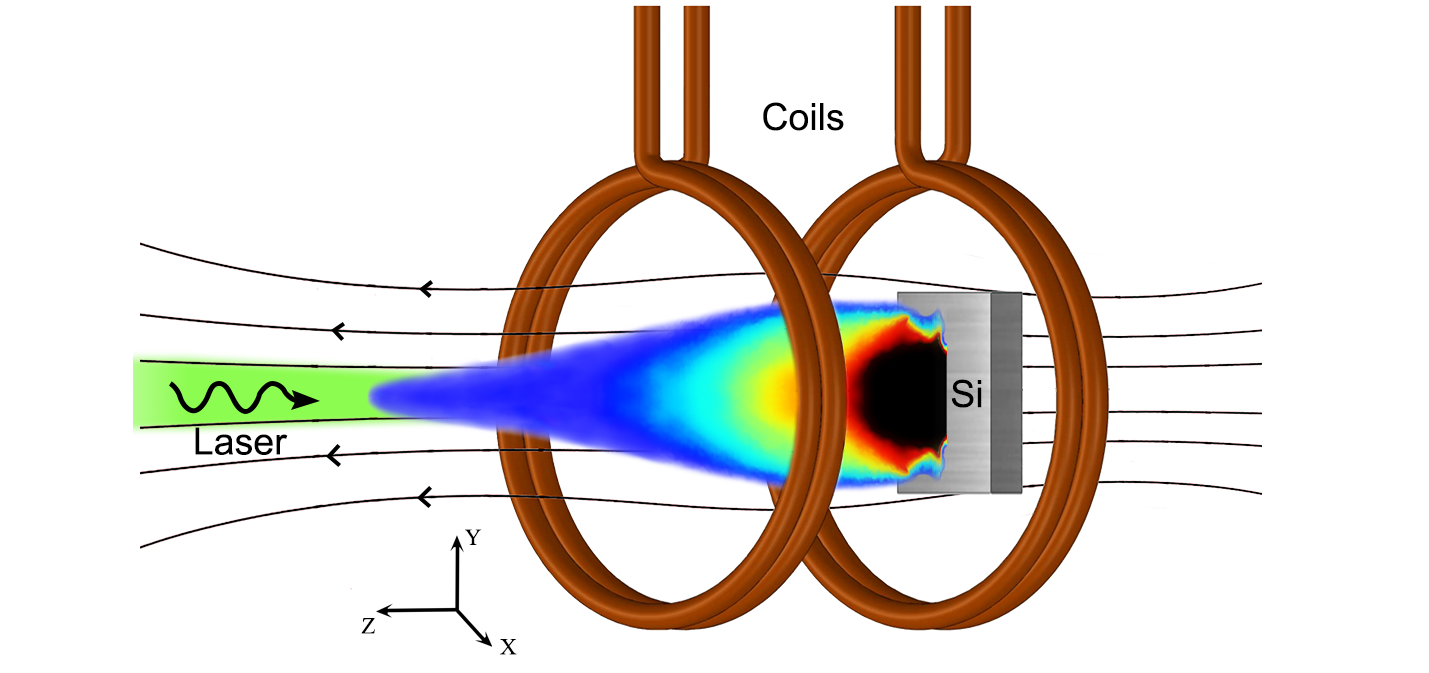}
\caption{\textbf{Schematic of the experimental setup of laser-produced scaled astrophysical outflow.} }\label{fig:1}
\end{figure}

Fig.~\ref{fig:1} illustrates the experimental setup. The laboratory explosive plasma flow mimicking  astrophysical outflows is produced by a  high-power nanosecond laser pulse ablating a planar solid Si target. 
A pair of current coils produces a quasi-uniform axial magnetic field $\bm{B}_{ext}$ of strength $\sim7$~T and duration $>1~\mu s$, well covering the plasma outflow region $\varnothing 1$~cm$\times 9$~mm, to mimic the poloidal magnetic field in ISM.
Partially ionized helium ambient gas in the vacuum chamber is used to mimic the ambient slow wind of ISM around the explosive astrophysical outflows. The thermal-to-magnetic pressure ratio  $\beta$ in ambient gas is much less than unity. The magnetic field, ambient gas, and outflow are thus tailorable, so that we can scale-model different astrophysical outflows.

The laboratory plasma flow is highly conductive, wide-angle quasi-spherical, and initially dominated by its kinetic energy with ram speed $>250$~km/s (Fig.~\ref{fig:2}B-i), similar to the astrophysical fast wind. A time span of 35~ns in the laboratory can thus be scaled to $2.37\times10^{18}$~ns (equivalent to $\sim75$ light years), similar to that of the YSOs and PNs  (for examples, HH212's NK1 and SK1 knots)~\cite{zinnecker1998symmetrically,bjerkeli2016resolved} , during which the outflow extends forward for over $2.36\times10^{11}$~km (or 1580~AU) 
. In a distance of $\sim$ 2.5~mm in the laboratory, the outflow is halted by the magnetic and/or ambient gas pressures, with its kinetic energy converted into magnetic and thermal energies. The Reynolds ($Re$), magnetic Reynolds ($Re_M$), and Peclet ($Pe$) numbers are all much larger than unity, so that the ideal MHD theory is applicable, ensuring its relevance as a scaled astrophysical system.

\section*{Results}\label{sec3}

\begin{figure}[h]
\centering
\includegraphics[width=1\textwidth]{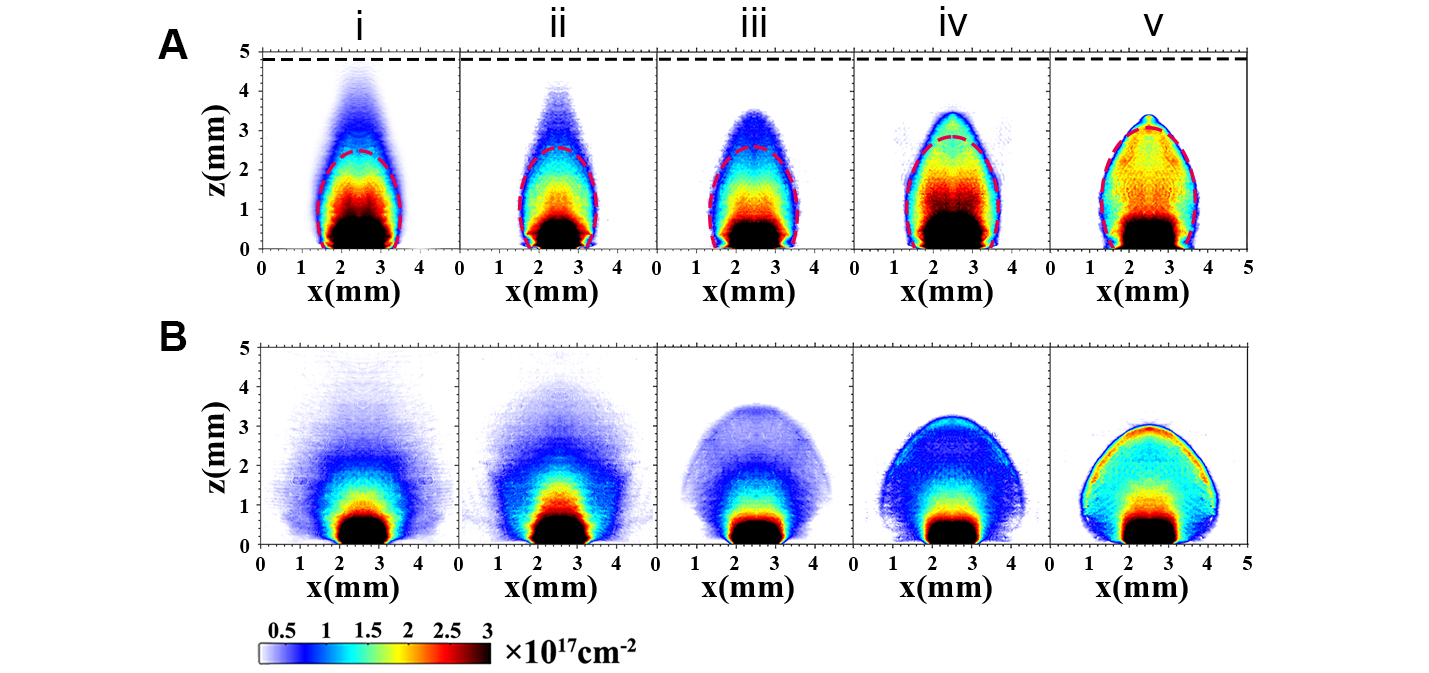}
\caption{\textbf{Laboratory observed morphologies of laser-driven scaled astrophysical outflow in ambient gas medium.} The maps show two cases with (panel \textbf{A}), and without (panel \textbf{B}) the applied 7 T axial magnetic field $\bm{B}_{ext}$, observed at 15 ns after the laser irradiation. The ambient gas pressure $P_{amb}$ and the external Alfvénic Mach number  $M_{e-a}$ in the columns \textbf{i} to \textbf{v} are $0.06 Pa $  and 0.014, $10 Pa $ and 0.18, $100 Pa $ and 0.57, $400 Pa $ and 1.14, $1000 Pa $ and 1.81, respectively. The corresponding experimental parameters are provided in~\ref{tab:S1}. The edges of the bubbles/cavities are indicted by dashed red curves. The horizontal dashed black lines at $z$ $\sim 4.8$ mm indicate the position of the current coils.}\label{fig:2}
\end{figure}

Typical morphologies of the laboratory plasma outflow in near-vacuum axial magnetic field ($P_{amb}<10^{-1}$~Pa) are shown in Fig.~\ref{fig:2}A-i. The axial magnetic field shapes the plasma flow to form a collimated jet with an aspect ratio of $\gtrsim4$, which later evolves to $>10$ (see also~\ref{fig:S4} in the Supplementary Materials). Both of our experiments and MHD simulations show that the initially wide-angle plasma flow, constricted and guided by the axial field, produces a converging elliptical bubble (or cavity) firstly. Subsequently, the expanding plasma flow refracts and slides along the wall of the elliptical bubble, finally forms the collimated jet at the apex of the bubble via a conical shock (see~ \ref{fig:S2}, \ref{fig:S4}, \ref{fig:S5}, and \ref{fig:S7} in the Supplementary Materials). This scenario is in full agreement with that in the available astrophysical and laboratory observation results ~\cite{albertazzi2014laboratory, ciardi2013astrophysics}.

\begin{figure}
	\centering
	\includegraphics[width=0.8\textwidth]{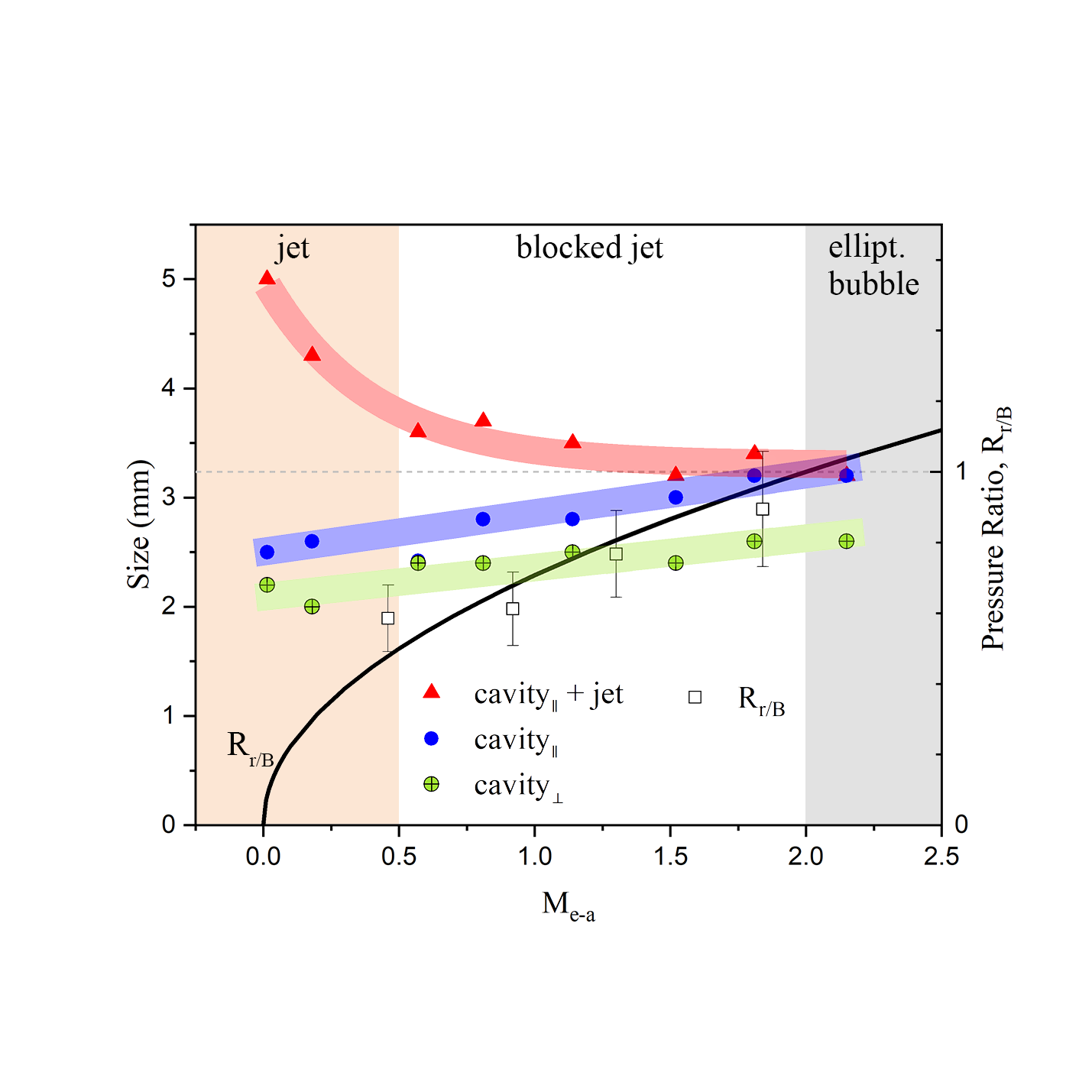}
	\caption{\textbf{Experimental measured bubble (or cavity) size and jet length versus the external Alfvénic Mach number $M_{e-a}$.} 
     Red triangles are for the combined cavity and jet lengths (longitudinal/axial). Blue circles and crossed green circles are for the longitudinal ($\parallel$) and transverse ($\perp$) sizes of the bubbles/cavities, respectively. Hollow squares are for the  ratio $R_{r-B}$ of the ram to magnetic pressure as obtained from the MHD simulations, which exhibits a function of $R_{r-B}\sim (M_A/2)^{1/2}$ denoted by the black curve. One can see that in the left (collimated jet) and center (blocked jet) regions the jet length (distance between red triangles and blue circles at a given $M_{e-a}$) decreases as $M_{e-a}$ increases. Moreover, the longitudinal size of the bubble/cavity is larger than the transverse size, which exhibits as an elliptical bubble. The corresponding experimental parameters are provided in~\ref{tab:S1} of the Supplementary Materials.} 
	\label{fig:3}
\end{figure}

The above discussed collimated jet formed in near-vacuum conditions will transition into a blocked jet or an elliptical bubble when introducing tenuous ambient gas into the axial magnetic field. Fig.~\ref{fig:2}A and Fig.~\ref{fig:3} show that as ambient gas in the vacuum chamber is increased, the length of the collimated jet decreases and an elliptical bubble begins to grow gradually. Fig.~\ref{fig:2}\textbf{A-iii} shows that at $P_{amb}=100$~Pa, or $\beta\sim1.7\times10^{-4}$, a clear elevated-density boundary emerges at the jet's head, similar to that observed in blocked astrophysical jets. Moreover, Fig.~\ref{fig:2}\textbf{A-v} shows that when $P_{amb}>1000$~Pa, or $\beta\sim6\times10^{-4}$, the jet disappears entirely, leaving only a roughly elliptical bubble.The elliptical bubble persists even at $P_{amb}=1400$~Pa and at later stage (see also~\ref{fig:S2},  in the Supplementary Materials), and have been observed in planetary nebulas (PNe)~\cite{bally2016protostellar,balick2002shapes,frank1999bipolar}. 

Our experimental and simulation results (Fig.~\ref{fig:3}) show that disappearance of the jet and appearance of the elliptical bubble occurs at $M_{e-a}$ $\sim$ 2, and the transformation from collimated jet to blocked jet occurs at $M_{e-a}$ $\sim$ 0.5, but the jet has no sharp high-density edge (The apparent boundary observed at 100~Pa is due to detection thresholds in our diagnostics). This gradual transition is accompanied by a continuous increase in the ambient gas pressure and $M_{e-a}$ in the range of $0<M_{e-a}<2$ and $M_{e-s}\gg1$.

The morphological transitions at $M_{e-a}$ $\sim$ 2 and 0.5 for $M_{e-s}\gg1$ from collimated jet to blocked jet and to elliptical bubble due to magnetic field and ambient gas are valid for a wide range of experimental parameters, and also (even quantitatively) consistent with our MHD simulation results (see~\ref{fig:S5} \& \ref{fig:S8}, and~\ref{tab:S2} \& \ref{tab:S3} in the Supplementary Materials).  The latter also include that for higher outflow velocities, magnetic fields, and gas pressures than that in the experiments. In general, we can see that higher plasma outflow velocities lead to larger bubbles, and stronger magnetic fields lead to smaller ones. Moreover, the bubble size $r$ follows the scaling $r\propto (V/B_{ext})^{2/3}$, which can be attributed to conversion of plasma-flow kinetic energy to ambient magnetic energy ~\cite{ciardi2013astrophysics}. The transitions and scaling laws of the outflow morphologies discussed here should be verifiable when more quantitative data from astronomical observations become available.

In the absence of the external magnetic field, the plasma outflow can be in the form of quasi-spherical wind or quasi-spherical bubble, depending on the pressure of the rarefied ambient gas (Fig.~\ref{fig:2}B and~\ref{fig:S2}). Fig.~\ref{fig:2}\textbf{B-iii} shows that the initially wide-angle quasi-spherical plasma flow acquires a halo  structure when $P_{amb}\sim100$~Pa. In contrast, Fig.~\ref{fig:2}\textbf{A-iii} shows that under the same $P_{amb}$, a blocked jet is formed if there is a sufficiently intense axial magnetic field. When $P_{amb}>1000$~Pa, the halos become more prominent, and the plasma flow becomes a spherical bubble, as can be seen in Fig.~\ref{fig:2}\textbf{B-v}. Moreover, ~\ref{fig:S3} and ~\ref{fig:S2} in the Supplementary Materials show that the longitudinal extents of the bubbles/halos are comparable with that of the bubble-plus-jet in the presence of the external magnetic field, but their transverse extent exceeds that of the latter. Our simulations indicate that halos (or bubbles) can appear if $M_s\geq 1$. In our experiments, $M_s\gg1$ holds even for very low $P_{amb}$. However, extremely low densities (say, $P_{amb}<100$~Pa) can make the halos undetectable.

\section*{Discussion}\label{sec4}
As a supersonic plasma flow enters and expands in the ambient gas, it drives an outer shock in ambient gas and a oppositely-propagating inner shock in the plasma flow itself, separated by a contact discontinuity (the interface between plasma flow and ambient gas). In the presence of an external axial magnetic field, competition (quantified by $M_{e-a}$ and  $M_{e-s}$) among the anisotropic magnetic pressures, isotropic ram and gas pressures in the downstream of the outer shock contributes to the confinement and shaping of the outflow, resulting in the asymmetric collimated jets, blocked jets, and elliptical bubbles. Without the constriction by the intense axial magnetic field, the outflow can be in the form of symmetric spherical bubbles and winds, depending on the competition (quantified by $M_{e-s}$) among isotropic ram and gas pressures. The magnetized outer shocks exhibit as either a single-layer (with magnetosonic Mach number $M_{a}>2$) or a switch on (with $1<M_{a}<2$) fast magnetosonic shocks (see~\ref{fig:S5} \& \ref{fig:S8} in the Supplementary Materials) ~\cite{de1998complex,de1999field,pogorelov2000nonevolutionary,ratkiewicz2003interstellar,somov2006plasma}. Both types degenerate into slow-moving sonic shock, similar to the characteristic of magnetically absent conditions, along the direction of magnetic field lines and jets.

In the presence of an intense axial magnetic field, such that $V_A\gg C_s$, our simulations reveal that the ram-to-magnetic pressure ratio $R_{r/B}$ acting on the contact discontinuity (the surface of plasma outflow) satisfies $R_{r-B}\sim (M_A/2)^{1/2}$ (see Fig.~\ref{fig:3} and~\ref{fig:S3} \&~\ref{fig:S6} \&~\ref{fig:S8}). At $M_{e-a}\sim 2$, the outer shock moves at the fast magnetosonic Mach number $M_a\sim1.5$, with magnetic and density compression ratio of ~$r\sim1.6$ from the Hugoniot relations ~\cite{somov2006plasma}
\begin{equation}
    \frac{1}{r}=\frac{1}{8}\left\{ \frac{5\beta}{2M_{a}^2} +1+\frac{5}{2M_{a}^2}+\left[\left(\frac{5\beta}{2M_{a}^2} +1+\frac{5}{2M_{a}^2}\right)^2+\frac{8}{M_{a}^2}\right]^{1/2} \right\}.
    \label{eq:1} % Use a logical label
\end{equation}
and $R_{r/B}\sim1$ at $M_{e-a}\sim 2$. For $M_{e-a}$ $>$ 2 with $R_{r/B}>1$, the ram pressure dominates, resulting in elliptical bubbles. In contrast, for $M_{e-a}<2$  with $R_{r/B}<1$, the anisotropic magnetic pressure dominates, resulting in jet. If $M_{e-a} < 2$ as well as $M_{e-s} < 1$, a well-collimated jet without obstruction results. In our experiments, the jet is always supersonic ($M_{e-s}>1$) and has a sonic shock at its front. The sonic shock is due to the ram pressure of the outflow, which impedes jet propagation, leads to  speed reduction and plasma accumulation, and finally shortens the jet. If the plasma outflow at the head of the jet exceeds $M_{e-a}\sim0.5$ (at which plasma accumulation beyond the detection threshold in our experiments), blocked jets can be formed. Since a sonic shock impedes the plasma flow, in the weak magnetic field limit $V_{a}\ll C_{s}$, transition from spherical wind to spherical bubble can also occur for $M_{e-s}>1$. However, similar to that of the blocked jet, laboratory and astronomical observability of this transition depends strongly on the local conditions.

\section*{Conclusion}\label{sec5}

~\ref{fig:S9} in the Supplementary Materials summarizes the existence conditions of the five basic astrophysical outflow morphologies, which are separated by the external Alfvénic and sonic Mach numbers $M_{e-a}=V/V_a$ and $M_{e-s}=V/C_s$. These Mach numbers characterize the \emph{relative magnitudes} of ram ($\rho V^2$), magnetic ($B^2/2\mu_0$), and thermal ($\rho C_{s}^2$) pressures against the outflow, where $\rho$ is the density of ISM. They are related via the expressions $M_{e-a}=\sqrt{\frac{1}{2} \frac{\rho V^2}{B^2/2\mu_0}}$ and $M_{e-s}=\sqrt{\frac{\rho V^2}{\rho C_{s}^2}}$. 

Five basic astrophysical outflow morphologies, namely jets, blocked jets, elliptical bubbles, spherical bubbles, and spherical winds, are reproduced in our scaled laboratory experiments and numerical simulations. Repeatability and controllability in the laboratory enable quantification of the conditions of the outflow morphologies. Our findings reveal that the external Alfvénic and sonic  Mach numbers $M_{e-a}$ and $M_{e-s}$, representing the relative magnitudes of ram, magnetic, and thermal pressures in the ISM, can uniquely determine the outflow morphology, with transitions occurring at $M_{e-a}\sim2$ and 0.5, and $M_{e-s}\sim1$. 

Coexistence of different outflow morphologies can lead to more complex structures. For example, the outer-to-inner nested bi-lobes and jets in the bipolar planetary nebula M2-9~\cite{livio2001twin} could be due to multiple outbursts, since plasma expansion and consequent rarefaction can reduce the outflow ram pressure, leading to transition from bi-lobes (when $M_{e-a}>2$) to jets (when $M_{e-a}<1$). Our results can thus provide a quantitative base for understanding the diverse outflow morphologies in different astrophysical environments ~\cite{bally2016protostellar,balick2002shapes,frank1999bipolar,bondarenko2017collisionless,suzuki2015bow,tang2025laboratory,fiuza2020electron,li2019collisionless}).

\section*{Methods}\label{sec6}

\textbf{Experimental setup.} Our experiments are conducted in a vacuum chamber using a nanosecond Nd:YAG laser at the University of Science and Technology of China. A 470~mJ laser beam of $2\times 10^{12}{\rm W/cm^2}$ intensity, 532~nm wavelength, and 9~ns pulse duration incidents normally on a 525~$\mu$m-thickness planar silicon target to produce a high temperature plasma (see~\ref{fig:S1}). The plasma is thermally launched from a region with the laser spot diameter of $\sim$ 40~$\mu$m, and finally forms a supersonic wide-angle plasma flow with 20-50 eV temperature, $\sim$ 6 ion charge state, and speed exceeding 250 km/s~\cite{tang2020observation,zhang2023influence,tang2018confinement}. 

The silicon target and laser-produced plasma flow were embedded in an external axial magnetic field, oriented along the normal direction of planar silicon target. A Current-carrying copper coil pairs produce a pulsed axial magnetic field of $\sim7$~T peak intensity, $>$ 1~$\mu$s duration (oscillation period is 10~$\mu$s), and $\varnothing$ 1~cm $\times$ 9~mm region, which is quasi-uniform for the plasma outflow that expands to roughly $\sim \varnothing$ 3~mm $\times$ 5~mm size in $\sim$ 100~ns.

Helium ambient gas of tunable $10^{-2}$~Pa to $10^3$~Pa pressure in the vacuum chamber is used to mimic the ISM that surrounding the astrophysical outflows, that is rarefied enough to ensure $\beta\ll$ 1 in ambient gas medium.

The profiles of the plasma flow are diagnosed with an 800~nm, 30~fs probe laser beam, which passes through the plasma and ambient gas along the surface of the planar silicon target, simultaneously producing the optical interferometry and the dark-field schlieren images~\cite{si2023digital,tang2025laboratory}, which are recorded by CCD cameras (Manta G-609B, Allied-Vision). The former measures the line-integrated electron density profile, providing the electron density maps after Abel inversion. The latter generates the discontinuity-surfaces maps which indicates the first spatial derivative of the line integrated electron density profile. The electron-density detection threshold is around $1\times10^{18}{\rm /cm^3}$. 

\textbf{MHD simulations.} The outflows are simulated using the 2D resistive MHD code ~\textsc{FLASH 4.6.2}~\cite{fryxell2000flash} with Eulerian radiation, tabulated EoS and opacity table from IONMIX~\cite{macfarlane1989ionmix}. The instantaneous diffusion-related coefficients such as that of the heat conductivity, viscosity, and magnetic diffusing are calculated using the Spitzer collision model~\cite{spitzer1962beginnings}. The 2D cylindrical simulation box is $0.8~cm \times 2.4~cm$, and the spatial resolution is 50~$\mu$m. A 300~$\mu$m diameter silicon or aluminum planar target is irradiated by a 1.064~$\mu$m laser. The target diameter is significantly larger than the focal-spot diameter of 100~$\mu$m. The incident laser, focused by a $F/10$ lens, incidents on the target at an angle of either 45° (mainly for tuning the ambient gas) or 0° (mainly for tuning the magnetic field). The laser intensity is maintained at $3.3\times 10^{13}{\rm W/cm^2}$. Free boundary conditions are used on all sides of the simulation box, allowing free exit of the plasmas and magnetic fields. The resulting density and magnetic field profiles are shown in ~\ref{fig:S5} \& ~\ref{fig:S8} and~\ref{tab:S2} \& \ref{tab:S3}. 

%\bibliography{RF}%
%%====================bbl============================%

%%====================bbl============================%

\section*{Acknowledgements}\label{sec7}
%\bmhead{Acknowledgements}
We thank the LULI team, Xue-ning Bai (Tsinghua University), Chin-Fei Li (Taiwan), Bai-fei Shen (Shanghai Normal University), Ru-xin Li (ShanghaiTech University), Tao Tao (University of Science and Technology of China) for their help in evaluating the results and useful discussions. This work is supported by the Controversial and Disruptive Projects of Chinese Academy of Sciences (Grant No. FGSDFX-0001); 
Strategic Priority Research Program of Chinese Academy of Sciences (Grant No. XDB16000000);
National Natural Science Foundation of China (Grant Nos. 12175230, 11775223, 12205298);
USTC Research Funds of the Double First-Class Initiative” (Grant No. YD2140002006).

\section*{Author contributions }\label{sec8}
Guang-yue Hu and Hui-bo Tang conceived and led this project. The experiments were carried out by Shun-yi Yang, Guang-yue Hu, Chao Xiong, Tian-yi Li, Hui-bo Tang, Shuo-ting Shao, Xiang Lv, Xue-cheng Li, and Chen Zhang. Numerical simulations were performed by Shun-yi Yang, Xue-cheng Li, and Hui-bo Tang. Additional theoretical support was provided by Ming-yang Yu. The paper was written by Guang-yue Hu, Ming-yang Yu and Shun-yi Yang with contributions from all the authors. 

\section*{ Competing interests }\label{sec9}
All authors declare they have no competing interests.

\newpage
\appendix
\setcounter{figure}{0}
\renewcommand{\thefigure}{Extended Data Figure \arabic{figure}} %
\renewcommand{\figurename}{}
\renewcommand{\thetable}{Extended Data Table \arabic{table}} %
\renewcommand{\tablename}{}
\section*{Supplementary information}\label{sec10}

\begin{figure}[h]
	\centering
	\includegraphics[width=0.9\textwidth]{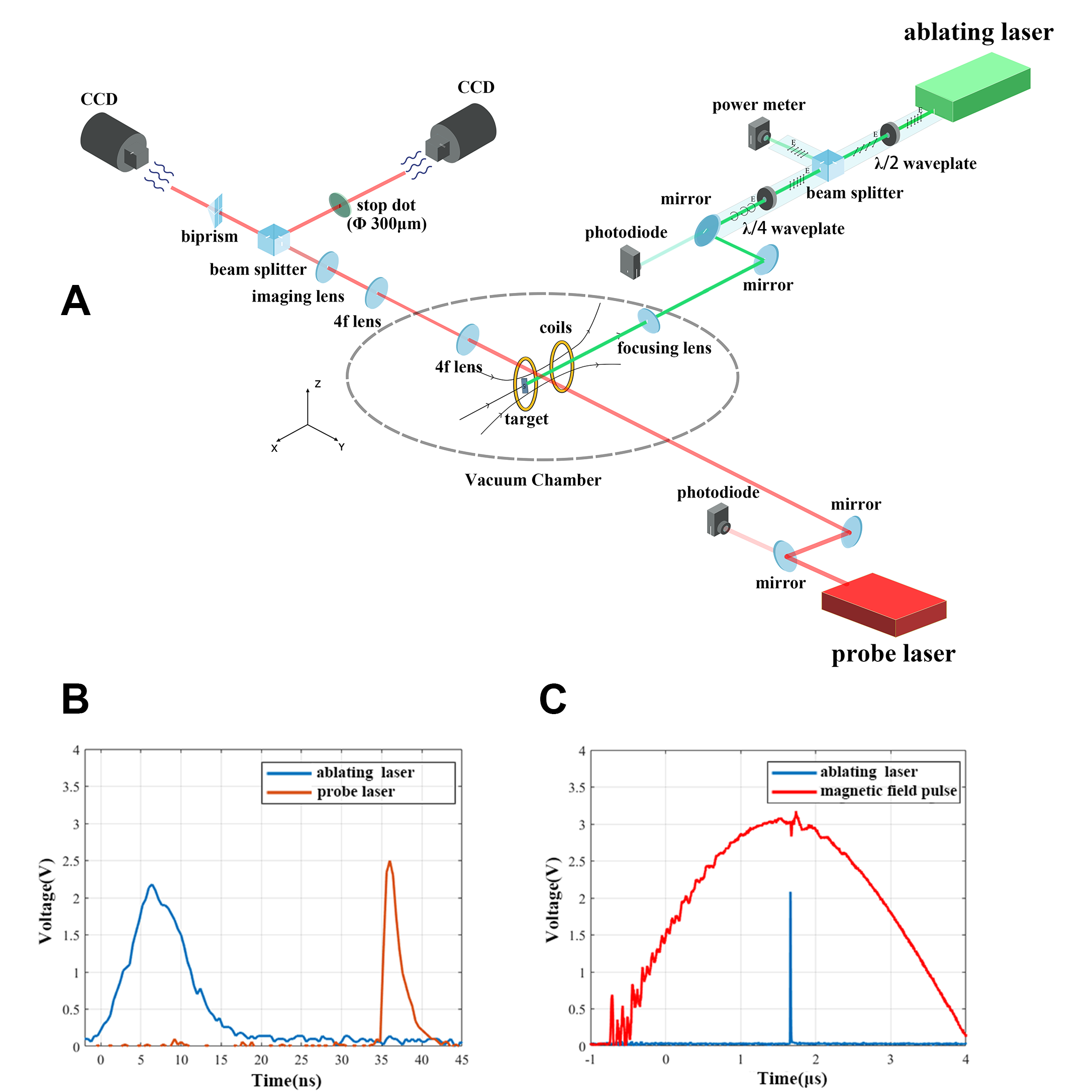} 
	\caption{\textbf{Schematic of the experimental setup.}
		(\textbf{A}) Nanosecond laser heats the planar silicon target surrounded by ambient gas and the current coils produce axial magnetic field. After passing through the plasma along target surface, a femtosecond-laser probe beam is split into two parts by a beam splitter. One part is for the optical interferometry. The other is for the dark-field schlieren, with a $\Phi$300~$\mu$m stop dot to block the zero-order light.  The temporal profile of pulsed magnetic field, the ablating laser beam, and fs probe laser beam are shown in (\textbf{B}, \textbf{C}), measured by photodiode (DET10A2, THORLABS) and home-made Rogowski coil.}
	\label{fig:S1}
\end{figure}

\newpage

\begin{figure}[h]
	\centering
	\includegraphics[width=0.9\textwidth]{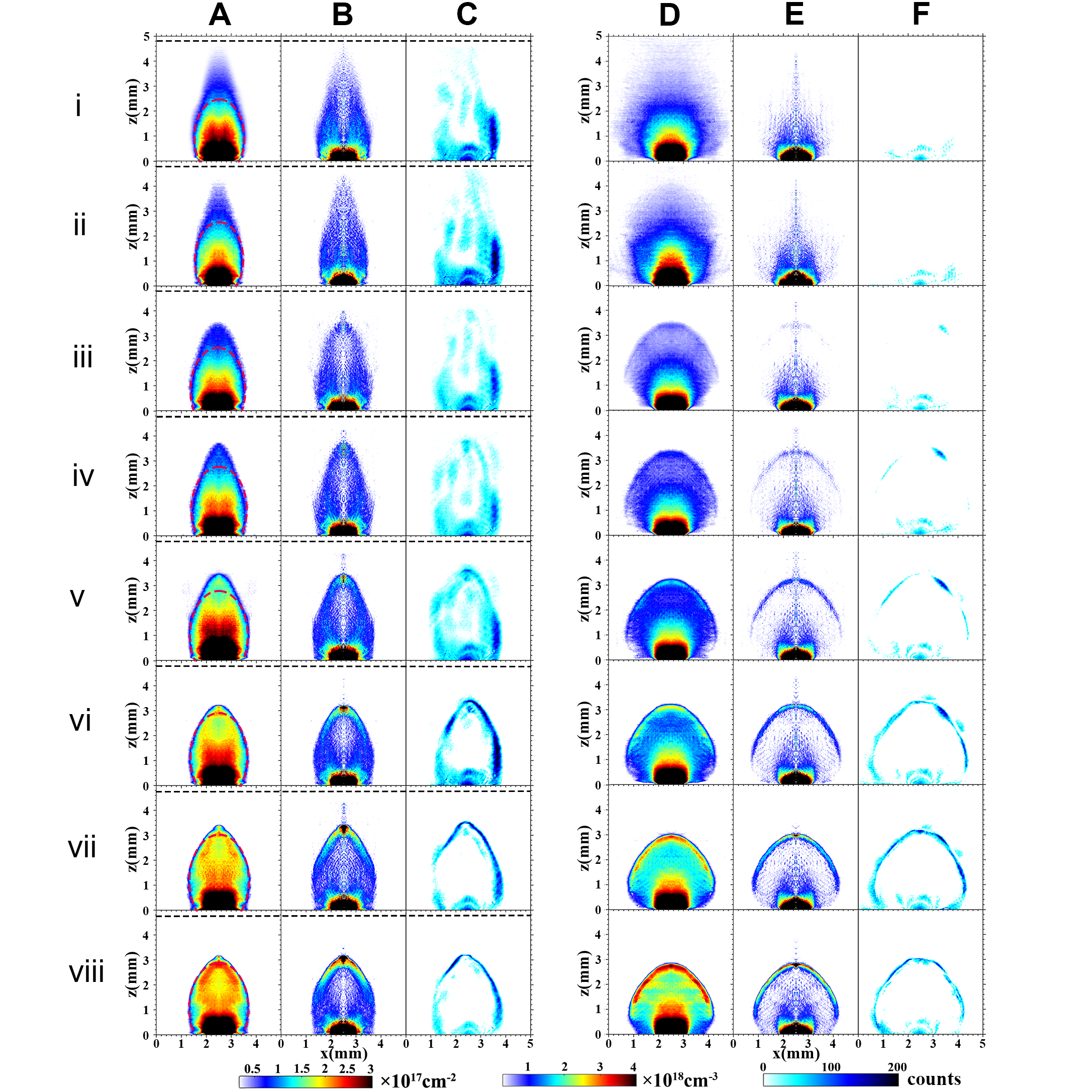} 
	\caption{\textbf{Experimental images of laser-driven scaled astrophysical outflow observed at 15~ns after laser irradiation of the target.}
		Line-integrated electron density distributions from optical interferometry (\textbf{A} and \textbf{D}), electron density maps from Abel inversion (\textbf{B} and \textbf{E}), dark-field schlieren fringes (\textbf{C} and \textbf{F}), for with (\textbf{A} to \textbf{C}) and without (\textbf{D} to \textbf{F}) the axial external magnetic field. The ambient gas pressures $P_{amb}$ and external Alfv\'enic Mach number $M_{e-a}$ in the rows \textbf{i} to \textbf{viii} are $<0.1$~Pa and $0.014$, $10$~Pa and $0.18$, $100$~Pa and $0.57$, $200$~Pa and $0.81$, $400$~Pa and $1.14$, $700$~Pa and $1.52$, $1000$~Pa and $1.81$, $1400$~Pa and $2.15$, respectively.
       The corresponding experimental parameters are given in~\ref{tab:S1}. The edges of the bubbles/cavities are indicated by dashed red curves with jet at the front in columns \textbf{A}.  The weak central linear patterns in columns \textbf{B} and \textbf{E} are numerical pseudo-data caused by the process of Abel inversion. 
       The horizontal dashed lines at $z\sim4800\mu$~m indicate the edge of the coils in columns \textbf{A} to \textbf{C}. }
	\label{fig:S2}
\end{figure}
In ~\ref{fig:S2}, jets appear in front of the elliptical bubble in the rows \textbf{A/B/C-i} to \textbf{A/B/C-iii}. One can also see in the rows \textbf{A/B/C-iv} to \textbf{A/B/C-viii} how shocks (very dense regions) are formed at the tips of the elliptical bubbles/cavities and the jets become blocked as the ambient gas pressure is increased. On the other hand, column \textbf{E} for the $B_{ext}=0$ cases shows that, as $P_{amb}$ is increased only spherical bubbles and shocks are formed without jets. In the images of dark-field schlieren fringes in columns \textbf{C}, The $1^{st}$ and $2^{nd}$ patterns around the head of the bubbles/cavity in \textbf{C-iv} to \textbf{C-vi} indicate the formation of blocked jet , while in \textbf{C-viii}, the $2^{nd}$ pattern disappears accompanying the jet vanishes entirely similar to the $B_{ext}=0$ cases in columns \textbf{F}.

\clearpage

\begin{figure}[h]
	\centering
	\includegraphics[width=0.9\textwidth]{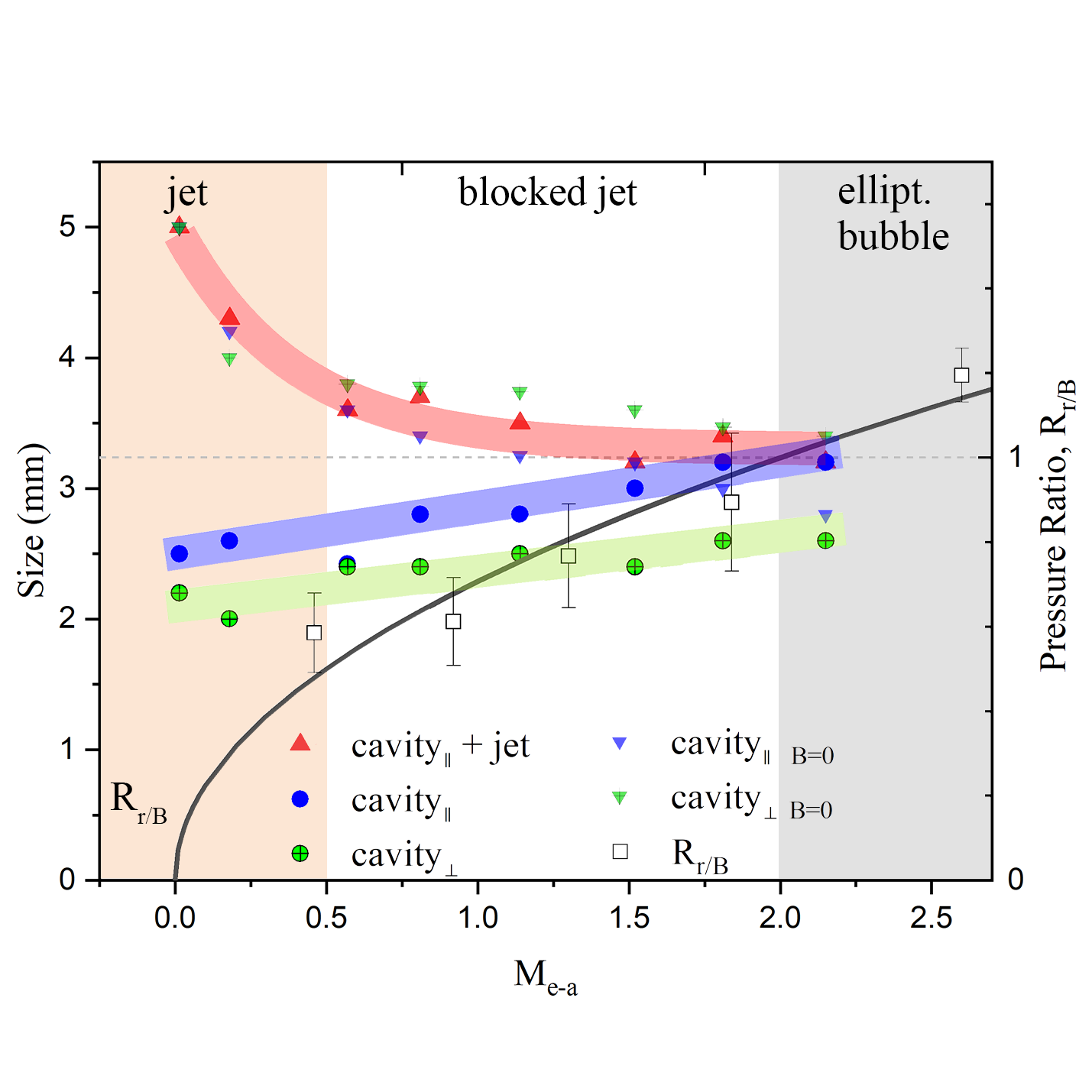} 
	\caption{\textbf{Bubble/cavity size and jet length versus the external Alfvénic Mach number ($M_{e-a}$) from the experiments.}
		Red triangles are for the combined bubble and jet longitudinal (axial) length. Blue circles are for the bubble longitudinal length ($\parallel$). Green circles with crosse are for the bubble transverse size (diameter, $\perp$). Hollow squares are for the ratio $R_{r-B}$ of the ram-to-magnetic pressures acting on the plasma flow as obtained from the simulations. Curve fitting (black curve) suggests the relation $R_{r-B}\sim\sqrt{M_A/2}$. For comparison, the corresponding bubble sizes (blue and green triangles) for the $B_{ext}=0$ cases are also shown. The experiment parameters are given in~\ref{tab:S1}.}
	\label{fig:S3}
\end{figure}

\clearpage

\begin{figure}[h] 
	\centering
	\includegraphics[width=0.9\textwidth]{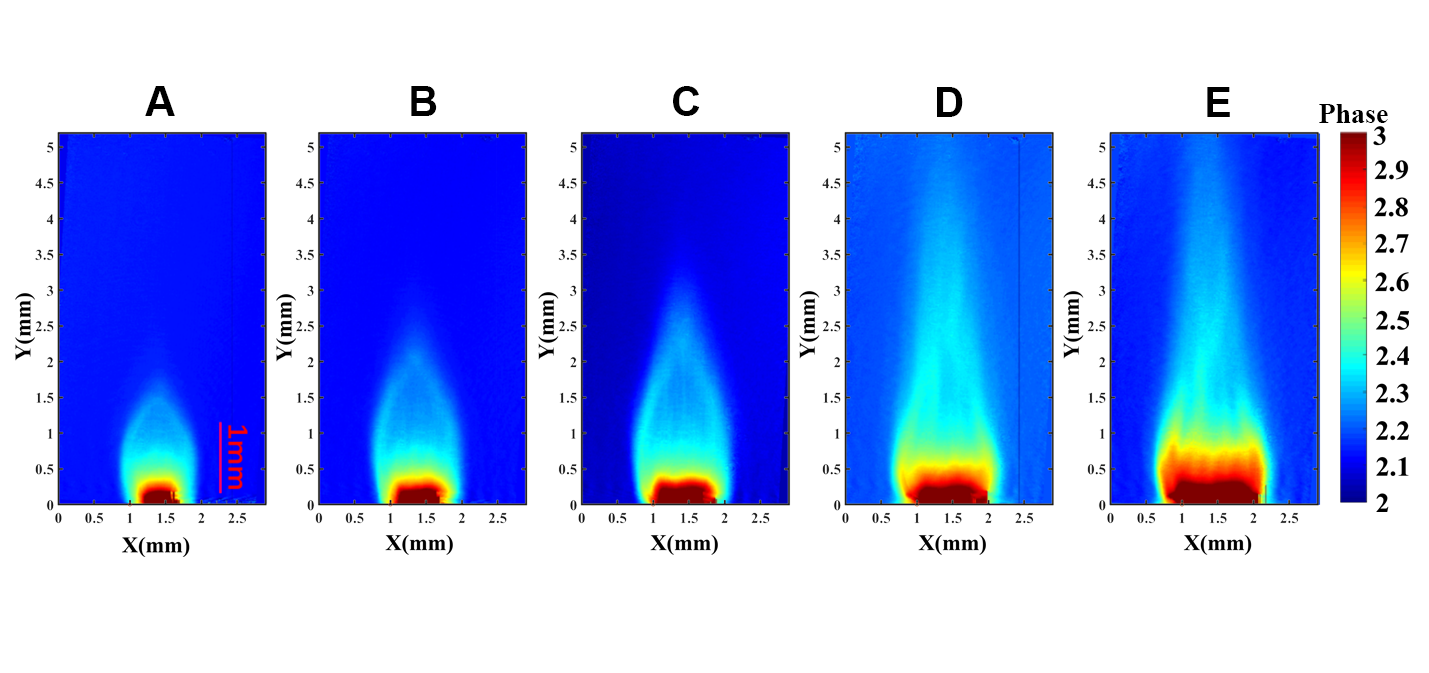} 
	\caption{\textbf{Experimental images of magnetized collimated jet observed at 3~ns (A), 5~ns (B), 7
    ~ns (C), 12~ns (D), 15~ns (E) after laser irradiation at $\bf 10^{-2}$~Pa near- vacuum condition.} Corresponding laser parameters are 600mJ energy, 532~nm wavelength, $2.7\times10^{12}$W/cm$^{2}$ laser intensity, 7~ns pulse duration. Magnetic field strength is 8~T.}
	\label{fig:S4}
\end{figure}

\clearpage

\begin{figure}[h] 
	\centering
	\includegraphics[width=1\textwidth]{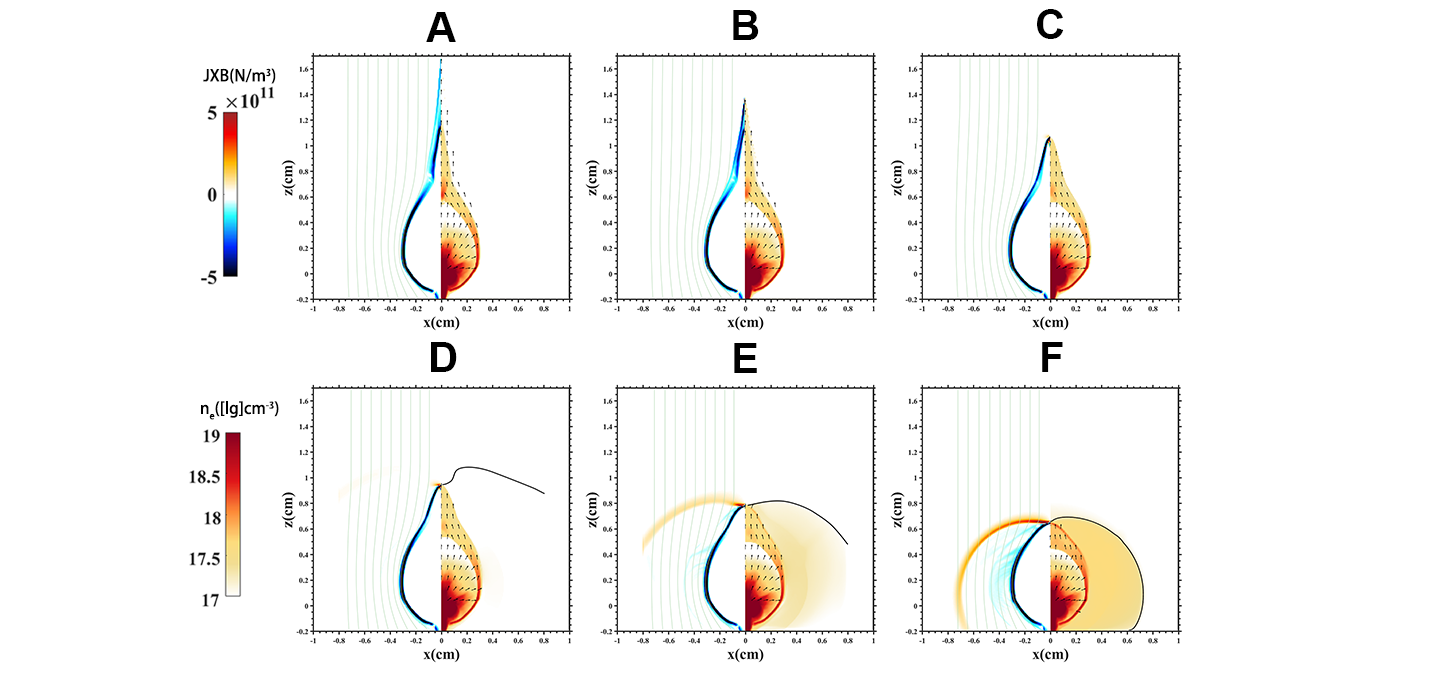} 
	\caption{\textbf{MHD simulation of laser-driven scaled astrophysical outflow with 20~T magnetic field and tuning ambient gas density. }  The left part is the magnetic field lines and $J\times B$ force, the right part is the plasma density. The ambient gas densities and the external Alfvénic Mach number $M_{e-a}$ are $1\times10^{-8}$g/cm$^{3} $ and 0.23 (\textbf{A}), $4\times10^{-8}$g/cm$^{3} $ and 0.46 (\textbf{B}), $1/6\times10^{-7}$g/cm$^{3}$ and 0.92 (\textbf{C}), $3.2\times10^{-7}$g/cm$^{3} $ and 1.3 (\textbf{D}), $6.4\times10^{-7}$g/cm$^{3} $ and 1.84 (\textbf{E}), $1.28\times10^{-6}$g/cm$^{3} $ and 2.6 (\textbf{F}) respectively. The maps are observed at 12~ns delay from laser onset with flat-top laser pulse of $3.3\times 10^{13}{\rm W/cm^2}$ intensity. The black curves label the surface of outer shock which exhibit as a switch on shock and a single-layer fast shock when $M_{e-a}<2$ and $M_{e-a}>2$ respectively.The corresponding simulation parameters are provided in~\ref{tab:S2}. }
	\label{fig:S5}
\end{figure}

\clearpage

\begin{figure}[h] 
	\centering
	\includegraphics[width=0.9\textwidth]{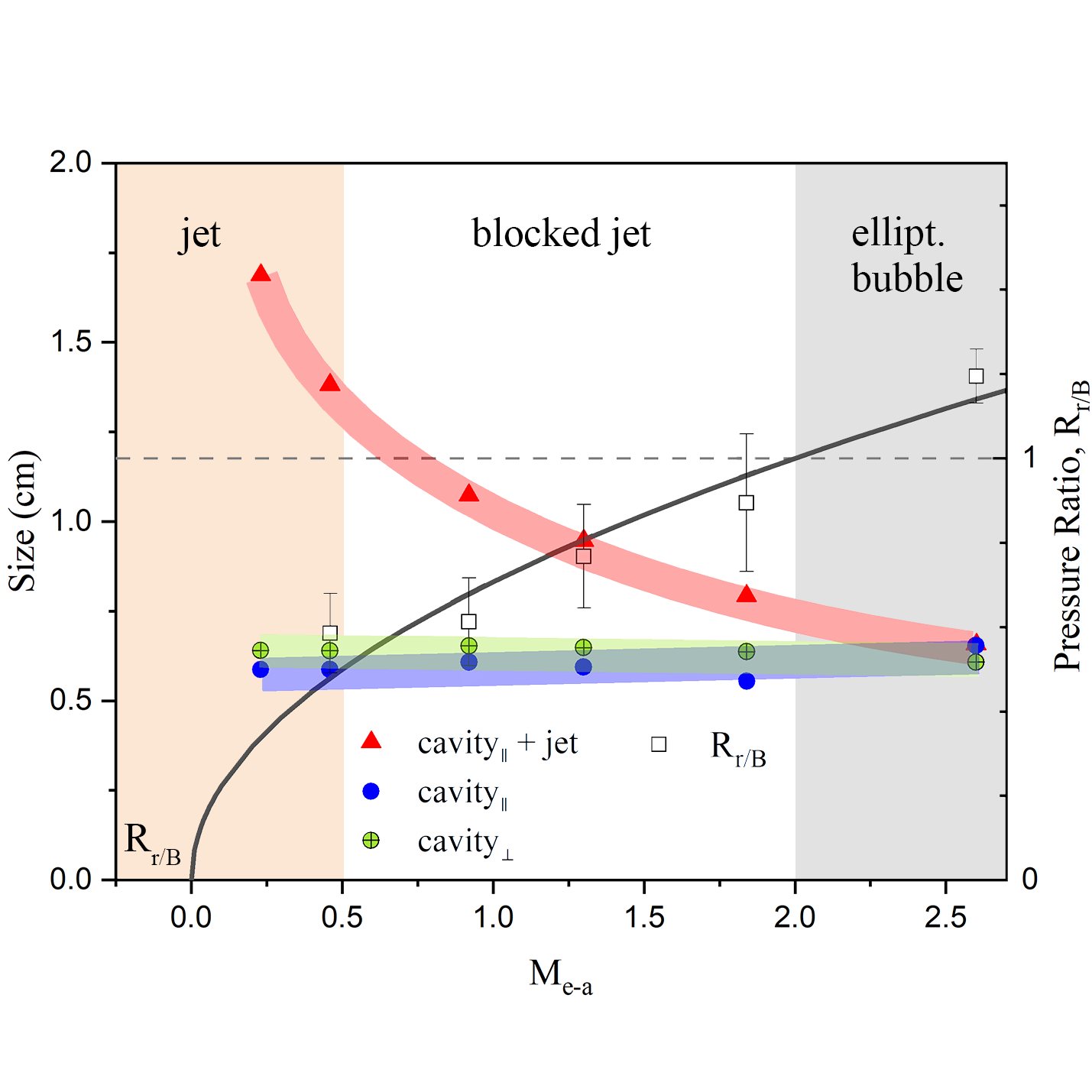} 
	\caption{\textbf{ MHD simulation results for bubble/cavity size and jet length versus the external Alfvénic Mach number ($M_{e-a}$) for $B_{ext}=20$~T and tuning ambient gas density.}
		Red triangles are for the combined longitudinal (axial) length of bubble-plus-jet . Blue circles are for the bubble longitudinal length ($\parallel$). Crossed green circles are for the bubble transverse size (diameter, $\perp$). Hollow squares are for the  ratio of the ram-to-magnetic pressure. The black curve is for the function $R_{r-B}=(M_A/2)^{1/2}$.The corresponding simulation parameters are provided in~\ref{tab:S2}.}
	\label{fig:S6}
\end{figure}

\clearpage

\begin{figure}[h] 
	\centering
	\includegraphics[width=1\textwidth]{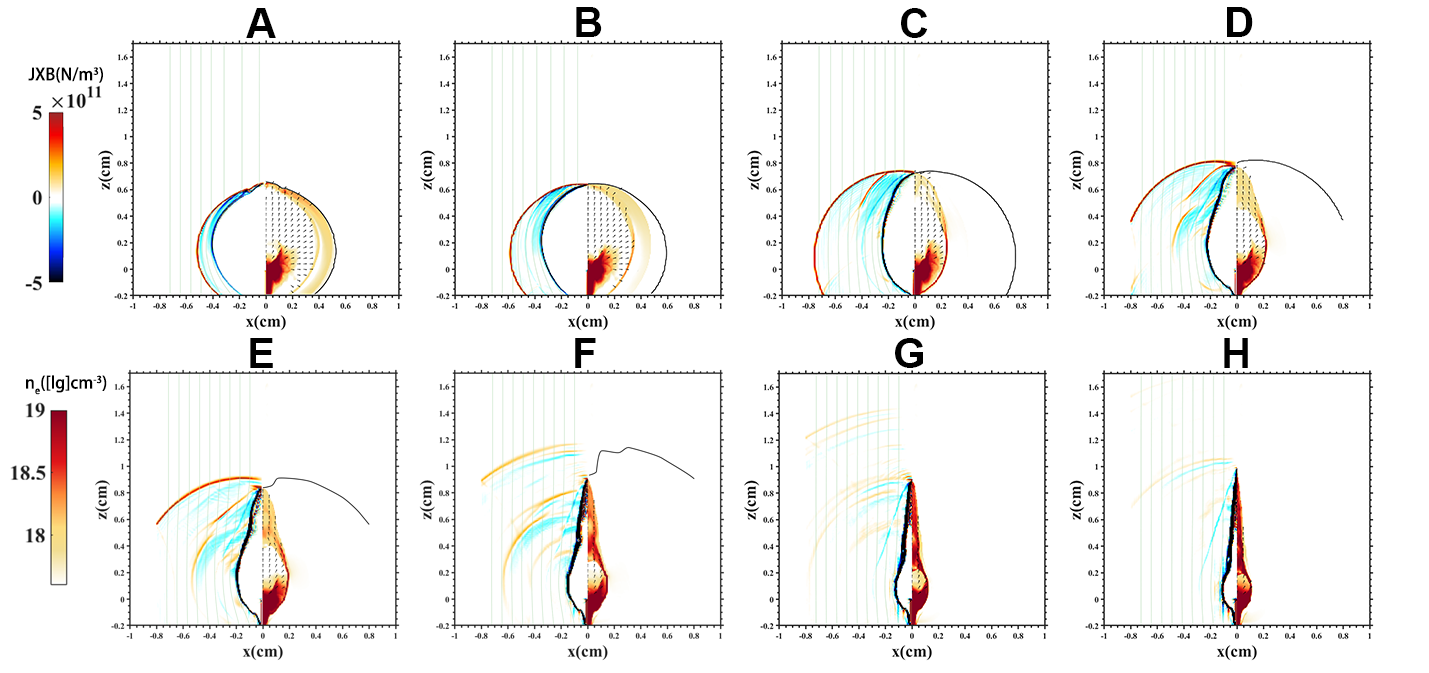} 
	\caption{\textbf{MHD simulation of laser-driven scaled astrophysical outflow with ambient gas density  $1.28\times10^{-6}$g/cm$^3$ and tailored magnetic field.}
		The left part of the outflow in the figures is for the magnetic field lines (green) and the $J\times B$ force (color bar), the right part is for the plasma density (color bar). The magnetic field intensities and external Alfvénic Mach number $ M_{e-a}$ are 5~T and 13.4 (\textbf{A}), 10~T and 6.7 (\textbf{B}), 20~T and 3.3 (\textbf{C}), 25~T and 2.7 (\textbf{D}), 30~T and 2.2 (\textbf{E}), 40~T and 1.7 (\textbf{F}), 50~T and 1.3 (\textbf{G}), 60~T and 1.1 (\textbf{H}), respectively. All results are for 12~ns after the laser impact on the target. The intensity of the flat-top laser pulse is $3.3\times 10^{13}{\rm W/cm^2}$ . Details of the simulation parameters are given in~\ref{tab:S3}.}
	\label{fig:S7}
\end{figure}

\clearpage

\begin{figure}[h] 
	\centering
	\includegraphics[width=0.9\textwidth]{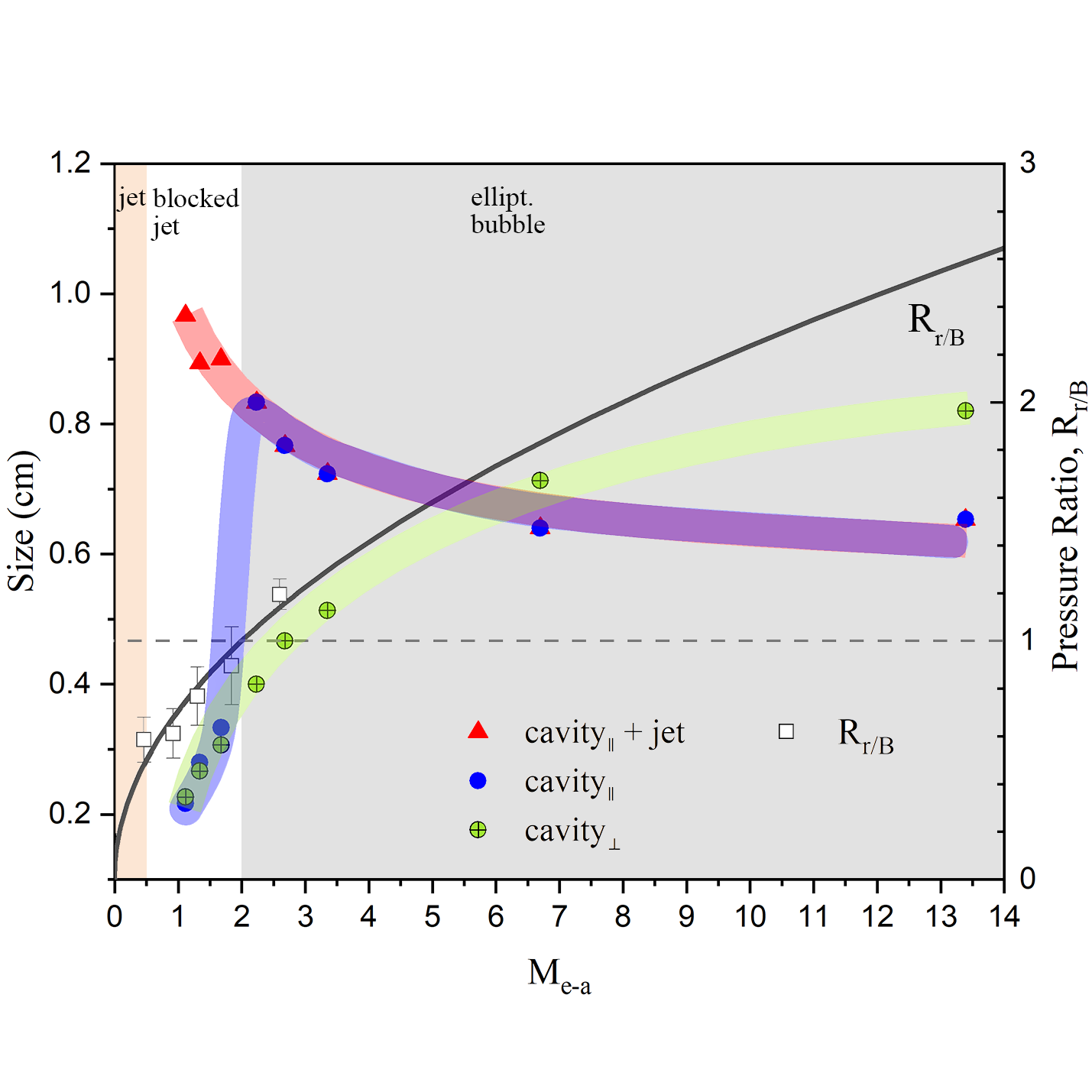} 
	\caption{\textbf{MHD simulation results for the bubble/cavity size and jet length versus the external Alfvénic Mach number $M_{e-a}$ for ambient gas density $1.28\times10^{-6}$g/cm$^3$ and tailored magnetic field.}
		Red triangles are for the combined bubble and jet length (axial). Blue circles are for the bubble longitudinal length (axial, $\parallel$). Crossed green circles are for the bubble transverse size (diameter, $\perp$). Hollow squares are for the ratio $R_{r/B}$ of ram-to-magnetic pressure. The black curve is for the function $R_{r-B}=(M_A/2)^{1/2}$. Details of the simulation parameters are given in~\ref{tab:S3}.}
	\label{fig:S8}
\end{figure}

\clearpage

\begin{figure}[h]
	\centering
	\includegraphics[width=0.85\textwidth]{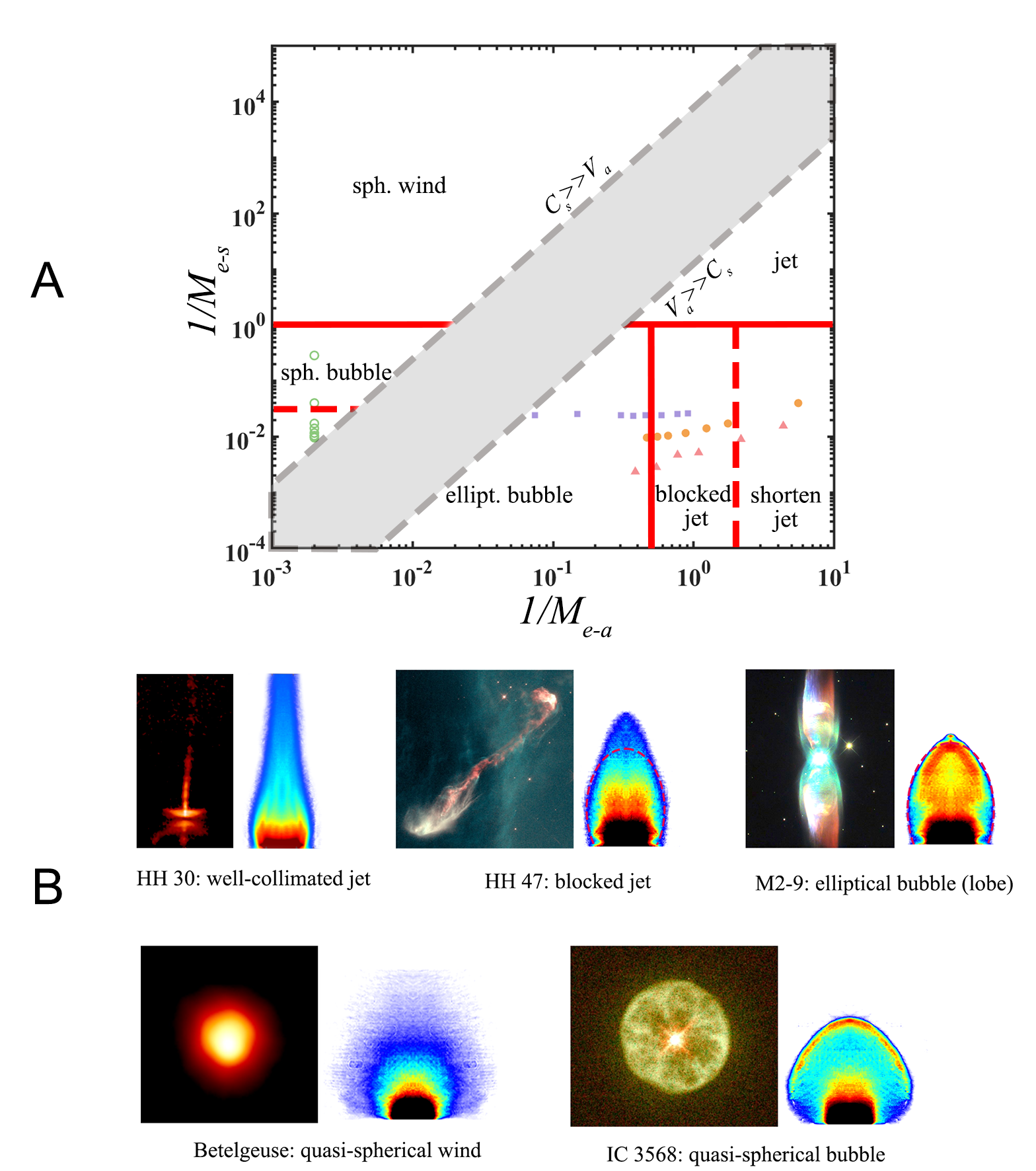} 
	\caption{\textbf{(A) Diagram of $1/M_{e-s}$ versus $1/M_{e-a}$ showing the existence conditions of the five basic astrophysical outflow morphologies in terms of the external Alfvénic and sonic Mach number.} Orange and green circles are from the experiments with and without the external magnetic field, respectively. Red triangles and purple squares are from the MHD simulations by varying the ambient gas density and external magnetic field, respectively. \textbf{(B) Common outflow morphologies in astronomical (left) and our laboratory (right) observations.} Note that the experimental initial plasma outflows settings are half-spherical, not the spherical initial outflows in astronomical observation.}
    \label{fig:S9}
\end{figure}

Note that $M_{e-a}$ and $M_{e-s}$ are related to the ratios of ram-to-magnetic pressure since $M_{e-a}=\sqrt{\frac{1}{2} \frac{\rho V^2}{B^2/2\mu_0}}$, and ram-to-thermal pressure since $M_{e-s}=\sqrt{\frac{\rho V^2}{\rho C^2}}$, respectively. 
Thus, a $1/M_{e-s}$ versus $1/M_{e-a}$ plot can be separated into two regions: the upper-left part ($V_A<C_s$) for the thermal-pressure dominated regime, and the lower-right part ($V_a > C_s$) the magnetic-pressure dominated regime. Here, we focus solely on two extreme cases ($V_A\ll C_s$ or $V_A\gg C_s$):

(1) In the magnetic-field-free, thermal-pressure-dominated regime $V_A\ll C_s$, the $M_{e-s} = 1$ line separates the spherical winds ($M_{e-s} < 1$) from the spherical bubbles ($M_{e-s} > 1$). Recall that observation of spherical bubbles may be constrained by detection thresholds (e.g., the red dashed horizontal line) in both experiments and astronomical observations..

(2) In the axial magnetic field dominated region $V_A \gg C_s$, the $M_{e-a} = 2$ curve separates the elliptical bubbles from the jets. 
Elliptic bubbles require both $M_{e-a}>2$ and $M_{e-s}>1$. Jets can be separated by the $M_{e-s} = 1$ line into two sub-regions: well-collimated jets (if $M_{e-a} < 2$ and $M_{e-s} < 1$), and blocked jets ($M_{e-a} < 2$ and $M_{e-s} > 1$). Blocked jets can be further separated (say, by the vertical red dashed line) by detection thresholds in the experiments and astronomical observations. For example, in our experiments, the detection threshold is near $M_{e-a}\sim 0.5$: observable blocked jets lie in the region $0.5<M_{e-a}<2$ and $M_{e-s}>1$, whereas in the region $0< M_{e-a} < 0.5$ and $M_{e-s} > 1$, only jet shortening, with no clear blockage, is observed.

\clearpage

\begin{sidewaystable}[h]
\caption{\textbf{Experimental parameters.}}\label{tab:S1}
\begin{tabular*}{\textheight}{@{\extracolsep\fill}cccccccc}
\toprule%
outflow velocity & gas pressure & magnetic field & $\beta$ & Alfven velocity & sonic velocity & $M_{e-a}$ & $M_{e-s}$ \\
km/s & Pa& T & ~ & km/s & km/s & ~ & ~ \\
\midrule
250 & 0.06 & 6.7 & $3.3\times10^{-9}-2.8\times10^{-5}$ & 17830 & 72 & 0.014 & 3.5 \\ 
250 & 9.1 & 6.7 & $5.6\times10^{-7}-9\times10^{-5}$ & 1381 & 10 & 0.18 & 25 \\ 
250 & 101 & 6.7 & $5.6\times10^{-6}-1.7\times10^{-4}$ & 437 & 4.3 & 0.57 & 58 \\ 
250 & 205 & 6.5 & $1.2\times10^{-5}-2.4\times10^{-4}$ & 308 & 3.5 & 0.81 & 71 \\ 
250 & 402 & 6.7 & $2.2\times10^{-5}-3\times10^{-4}$ & 218 & 2.9 & 1.14 & 86 \\ 
250 & 702 & 6.7 & $3.9\times10^{-5}-4.3\times10^{-4}$ & 165 & 2.6 & 1.52 & 96 \\ 
250 & 1000 & 6.5 & $5.9\times10^{-5}-6\times10^{-4}$ & 138 & 2.5 & 1.81 & 100 \\ 
250 & 1399 & 6.7 & $7.8\times10^{-5}-7.3\times10^{-4}$ & 117 & 2.4 & 2.15 & 104 \\ 
250 & 0.1 & ~ & ~ & ~ & 72 & ~ & 3.5 \\ 
250 & 13 & ~ & ~ & ~ & 10 & ~ & 25 \\ 
250 & 110 & ~ & ~ & ~ & 4.3 & ~ & 58 \\ 
250 & 203 & ~ & ~ & ~ & 3.5 & ~ & 71 \\ 
250 & 403 & ~ & ~ & ~ & 2.9 & ~ & 86 \\ 
250 & 703 & ~ & ~ & ~ & 2.6 & ~ & 96 \\ 
250 & 1001 & ~ & ~ & ~ & 2.5 & ~ & 100 \\ 
250 & 1389 & ~ & ~ & ~ & 2.4 & ~ & 104 \\
\botrule
\end{tabular*}
\footnotetext{Including the external Alfvénic and sonic Mach numbers \textit{Me-a} and \textit{Me-s}, outflow speed, ambient gas pressure, magnetic field intensity, $\beta$ of ambient gas, Alfven velocity, sonic velocity.}
\end{sidewaystable}

\begin{sidewaystable}[h]
\caption{\textbf{Experimental parameters.}}\label{tab:S2}
\begin{tabular*}{\textheight}{@{\extracolsep\fill}ccccccccc}
\toprule%
outflow velocity & gas density & magnetic field & $\beta$ & Alfven velocity & sonic velocity & $M_{e-a}$ & $M_{e-s}$ & $R_{r/B}$ \\
km/s & g/cm$^{3}$ & T & ~ & km/s & km/s & ~ & ~ & ~\\
\midrule
1272 & $1\times10^{-8}$ & 20 & $3.9\times10^{-8}-3\times10^{-4}$ & 5642 & 20 & 0.23 & 64 & ~ \\ 
1272 & $4\times10^{-8}$ & 20 & $1.6\times10^{-7}-4.6\times10^{-4}$ & 2821 & 11.5 & 0.46 & 111 & 0.5857 \\ 
1272 & $1.6\times10^{-7}$ & 20 & $6.2\times10^{-7}-6\times10^{-4}$ & 1410 & 6.6 & 0.92 & 193 & 0.6123 \\ 
1272 & $3.2\times10^{-7}$ & 20 & $1.2\times10^{-6}-0.001$ & 997 & 6 & 1.3 & 212 & 0.7677 \\ 
1272 & $6.4\times10^{-7}$ & 20 & $2.5\times10^{-6}-0.0014$ & 705 & 3.6 & 1.84 & 353 & 0.88946 \\ 
1272 & $1.28\times10^{-6}$ & 20 & $5\times10^{-6}-0.002$ & 499 & 3 & 2.6 & 424 & 1.195 \\
\botrule
\end{tabular*}
\footnotetext{Including outflow speed, ambient gas density, magnetic field intensity, $\beta$ of ambient gas, Alfven velocity, sonic velocity, \textit{Me-a}, \textit{Me-s}, the ratio of ram-to-magnetic pressure $R_{r/B}$.}
\end{sidewaystable}

\begin{sidewaystable}[h]
\caption{\textbf{Experimental parameters.}}\label{tab:S3}
\begin{tabular*}{\textheight}{@{\extracolsep\fill}cccccccc}
\toprule%
outflow velocity & gas density & magnetic field & $\beta$ & Alfven velocity & sonic velocity & $M_{e-a}$ & $M_{e-s}$ \\
km/s & g/cm$^{3}$ & T & ~ & km/s & km/s & ~ & ~ \\
\midrule
1670 & $1.28\times10^{-6}$ & 5 & $8\times10^{-5}-0.06$ & 125 & 40.9 & 13.4 & 40.8 \\ 
1670 & $1.28\times10^{-6}$ & 10 & $2\times10^{-5}-0.018$ & 249 & 42.33 & 6.7 & 39.5 \\ 
1670 & $1.28\times10^{-6}$ & 20 & $5\times10^{-6}-0.004$ & 499 & 40.35 & 3.3 & 41.4 \\ 
1670 & $1.28\times10^{-6}$ & 25 & $3.2\times10^{-6}-0.0025$ & 623 & 40 & 2.7 & 41.8 \\ 
1670 & $1.28\times10^{-6}$ & 30 & $2.2\times10^{-6}-0.0018$ & 748 & 40.35 & 2.2 & 41.4 \\ 
1670 & $1.28\times10^{-6}$ & 40 & $1.2\times10^{-6}-0.001$ & 997 & 40.79 & 1.7 & 40.9 \\
1670 & $1.28\times10^{-6}$ & 50 & $8\times10^{-7}-6\times10^{-4}$ & 1247 & 42.87 & 1.3 & 39 \\
1670 & $1.28\times10^{-6}$ & 60 & $5.5\times10^{-7}-4\times10^{-4}$ & 1496 & 43.4 & 1.1 & 38.5 \\
\botrule
\end{tabular*}
\footnotetext{Including outflow speed, ambient gas density, magnetic field intensity, $\beta$ of ambient gas, Alfven velocity, sonic velocity, \textit{Me-a}, \textit{Me-s}, the ratio of ram-to-magnetic pressure $R_{r/B}$.}
\end{sidewaystable}

\newpage
\end{document}